\documentclass[aps,prl,twocolumn,groupedaddress,showpacs]{revtex4}
\usepackage{amsmath}
\usepackage{graphicx}

\begin{document}

\title{Crowding induced clustering under confinement}

\author{Remy Kusters and Cornelis Storm}
\affiliation{Department of Applied Physics and
Institute for Complex Molecular Systems, Eindhoven University of
Technology, P. O. Box 513, NL-5600 MB Eindhoven, The Netherlands}

\date{\today}

\begin{abstract}

We present Langevin dynamics simulations that study the collective behavior of driven particles embedded in a densely packed background consisting of passive particles. Depending on the driving force, the densities of driven and passive particles, and the temperature we observe a dynamical phase separation of the driven particles, which cluster together in tight bands. We explore the mechanisms that drive this cluster formation, and determine the critical conditions for such phase separation. A simple physical picture explains the formation and subsequent growth of a jammed zone developing in front of the driven cluster. The model correctly captures the observed scaling with time. We analyze the implications of this clustering transition for the driven transport in dense particulate flows, which due to a non-monotonic dependence on the applied driving force is not straightforwardly optimized. We provide proof-of-concept for a direct application of the clustering effect, and propose a 'colloidal chromatograph'; a setup that permits the separation of colloids by mass or size.

\end{abstract}

\pacs{64.75.Xc; 83.10.Mj; 83.50.Ha; 83.80.Hj}

\maketitle

In driven flows, particles are generally not alike. The collective motion in mixtures of driven and nondriven, or of differentially driven, particles has been the the focus of numerous studies. Examples arise in flows through random media \cite{Watson1996, Watson1997}, in lattice gases of oppositely charged particles under uniform external driving \cite{Schmittmann1992, Vilfan1994} and in mixtures of driven and non-driven colloids \cite{Reichhardt2006}. Mixtures of oppositely charged colloids driven by an external field  \cite{Glanz2012, Sutterlin2009, Rex2007} resemble situations encountered in pedestrian flows \cite{helbing1999optimal,evers2011} that obey similar basic laws, and show similar behaviors. Yet another way to realize differential driving is to have mixtures of differently sediment under the influence of gravity \cite{batchelor1982sedimentation,wedlock1990sedimentation,al1992sedimentation}. In each of these cases, heterogeneous driving - i.e., different forces on different subsets of member particles - gives rise to nontrivial collective behavior, and prompts some degree of spatial nonunifomity, either in the distribution of velocities or in the distribution of the particles themselves. The behavior becomes even richer when such systems are considered at high volume fractions. A common phenomenon in such systems is the occurence of {\em aggregation} of the driven particles. In the case of oppositely driven subsets, this generically manifests itself as a laning transition, where the driven particles organize spontaneously into filamentous aggregates, aligned along the direction of the driving. In mixtures of driven and nondriven particles, one tends to find small clusters in the bulk, a phenomenon that has been interpreted in terms of an effective attraction mediated by the surrounding, nondriven particles \cite{Reichhardt2006}.

\begin{figure}
\centering
\includegraphics[scale=0.73]{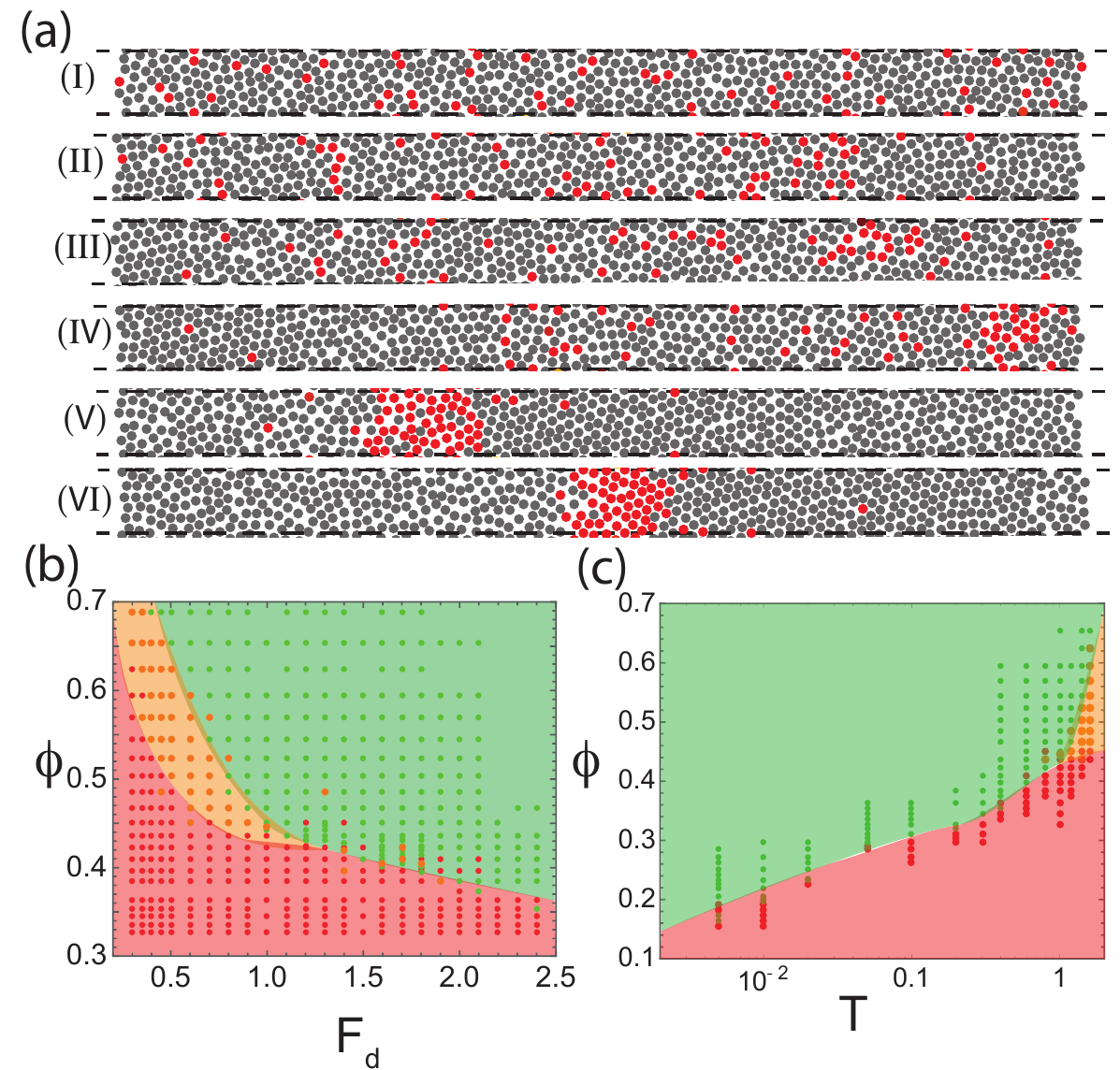}
\caption{(a) Snapshots of the clustering of the driven particles (Red indicates a driven particle). The initial configurations consists of $N_d=50$ driven and $N_p= 450$ passive partcles. (b) and (c) are the phase diagrams depending density $\phi$ and (b) Driving Force: $F_d$ ($T=1$) and (c) Temperature ($F=1$). The colors indicate the assymptotic state of the system is clustered $\sigma > 0.9$ (green), partial clustering  $0.9 > \sigma > 0.4$(orange) or no significant clustering $\sigma < 0.4$ (red). (b) and (c) contain $N_p=900$, $N_d=100$.\label{figure0}}
\end{figure} 

In this Letter, we ask the question what happens {\em after} this initial aggregation stage in confined systems. In many of the larger, bulk systems small clusters do not interact, and even when they do they do not encounter each other to any significant extent. When we confine our systems, we find that doing so prompts an unexpected nonlinear, late stage regime in the spatial organization of driven/nondriven mixtures: Our central finding is that, determined by the driving force, density, composition and temperature the driven particles very generically form a wall-to-wall pluglike cluster, phase separating from the passive particles. This late-stage behavior is very different from laning, as it involves organization perpendicular to the direction of driving rather than parallel to it. It is similarly distinct from bulk behavior, as all driven particles in a confined system will - under the right conditions - come together and. The appearance of this system-spanning aggregate has profound consequences for the further evolution of these systems: we show that it will completely dominate the global transport properties of the driven/nondriven mixture.

The basic effect is simple enough: As anyone who has ever attempted to guide a group of people through an otherwise stationary crowd can attest, those in front do all the work. The driven particles, on average, collide with more passive particles in front than at their rear. The momentum loss slows them down, but accelerates the passive particles in front. Further ahead of the driven particle, the average velocity is lower because the excess momentum is dissipated. As a result, there is a velocity gradient ahead of the particle which implies the accumulation of particles. This is indeed what is seen: a denser region in front of a driven particle develops. The excess particles in this denser zone must come from somewhere, and it cannot be the region ahead because this has not yet been affected by the driven particle. Therefore, the denser zone in front must be compensated for by a rarified zone in the wake of the driven particle.  Other driven particles moving in this wake are slowed down less by collisions.  This allows those in the rear to catch up to the frontmost particles. In bulk systems, this leads to small clusters of driven particles. In confined  systems, however, under the right dynamical conditions, the aggregation continues, ultimately resulting in  a stable cluster of driven particles, pushing a growing zone of jammed particles in front of it. 

Using the molecular dynamics package LAMMPS \cite{plimpton1995}, we perform Langevin dynamics simulations with a damping time of $\tau = 0.1$ Lennard-Jones time units. In the remainder of this Letter, all results are presented in Lennard-Jones units - allowing us to focus on the basic principles underlying the observed behavior. After an initial equilibration run with a soft repulsive potential between all particles, the interaction potential is replaced by the repulsive part of a 6-12 Lennard-Jonnes potential, where we cut off the potential at one Lennard-Jones length unit. We have verified that the results presented here are qualitatively independent of the precise choice of interaction potential (for details, we refer to the supplemental information). A constant force $F_d$ is exerted on each of the driven particles, superimposed upon their Brownian motion. 

\begin{figure*}
\centering
\includegraphics[scale=1.8]{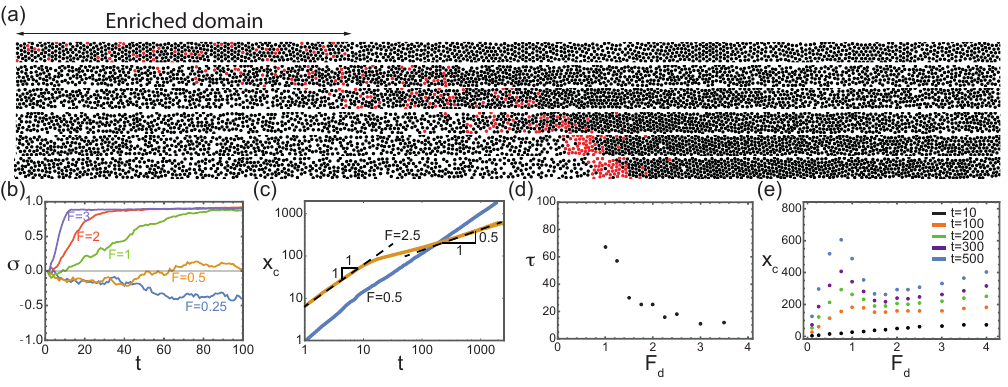}
\caption{(a) Snapshots of the clustering of an enriched region of containing 50 driven particles radomely distributed in a subdomain of size $-50 < z < 50$  (Here we show a fraction of the system, total system size is (length $l=4640$, width $w=6$ and $N_c=19950$). (b) $\sigma$ as function of time for various forces, where within this timeframe no clustering occurs for $F=0.25, 0.5$. Note that $\sigma$ becomes negative since we started with a enriched region of particles. (c) Time-evolution of the center of mass of the driven particles $x_d$ indicating the two distict regimes. (d) Time to reach a cluster $\tau$ as function of $F_d$. $x_d$ as function of $F_d$ for various time intervals. (e) Travelled distance of the center of mass as function of $F_d$. \label{fig:3}}
\end{figure*} 

\paragraph{Cluster mechanism.} We illustrate the phase separation in a two-dimensional channel with periodic boundaries at the four edges of the system (See Fig. \ref{figure0} (a)). We have performed simulations in two and three dimensions, for both doubly periodic and fixed walls, and have in each case found completely analogous behavior ( See Supporting Information). for clarity, we will present only the 2D doubly periodic images here but the phenomenon is very robust. The length of the system, which is the direction of forcing, is significantly larger than the width of the channel (length $l=100$ and width $w=6$, in LJ units). The system is initialized such that the driven particles (of which there are $N_d=50$) are homogeneously distributed through the disordered background of $N_c = 450$ Brownian particles (Stage I). The initial aggregation of particles (Stage II and III) is induced by cooperative behavior of the driven particles in a background of Brownian particles, an effect also seen in \cite{Habdas2004, Reichhardt2006}. Under our confined conditions however, and at sufficiently high driving force and overall density, however, these small clusters can continue to grow as the density of the passive particles rises in front of them (Stages V and VI). In these final stages of the evolution of this system - only seen under confinement - very generically, a single large plug develops. As it pushes against the particles in front, it densifies them and we can see the front edge of the compressed zone extending further and further into the system.

This is the general scenario: cluster initiation, maturation and subsequent densification of the passive particles ahead of it. In the following, we determine the critical conditions for cluster formation and then discuss the later stage dynamics of the densificaton front. In particular, we show that there is a growing lengthscale $\lambda$, associated with the size of the compressed region, in front of the cluster, and a characteristic velocity of the driven particles $\bar{v}$, similar to those observed in the dynamic jamming fronts induced by uni-axial compression of a nearly jammed particle packing \cite{Waitukaitis2013}.

\paragraph{Critical conditions for cluster formation.}
Whether or not the cluster forms depends on composition ($N_d$ and $N_c$), driving force $F_d$, density of the background $\phi$ and temperature $T$. To quantify this we measure the variance in z-position (parallel to the driving axis) of the driven particles, normalized to that of the initial configuration, and construct the following parameter: $ \sigma(t) = 1- \sum_{N_d} \left(x_i(t) - \bar{x}(t) \right)^2/\sum_{N_p} \left(x_i(0) - \bar{x}(0) \right)^2$, which approaches unity if a cluster is random closed packed and zero when there is no correlation present between the driven particles. In Fig. \ref{figure0} (b) and (c) we evaluate the asymptotic value of $\sigma$ as function of densities $\phi$ and (b) driving force $F_d$ ($T=1$) and (c) temperature $T$ ($F_d=1$) for a composition of $N_d=100$ and $N_c=900$. We use the orde parameter $\sigma$ to distinguish various degrees of clustering of the system: fully clustered when $\sigma > 0.9$, partially clustered when ($0.9> \sigma >0.4$) and not significantly clustered when $\sigma < 0.4$. Qualitatively we find that at high $F_d$ or low $T$, clustering occurs at lower $\phi$. At low $F_d$ and high $T$ we generally observe intermittent clustering: a region with high density of driven particles is generally observed to form and dissolve repeatedly. In this state, the thermal motion of both the passive and the driven particles is sufficient to prevent the formation of a single stable compact cluster.

\begin{figure}
\centering
\includegraphics[scale=1.8]{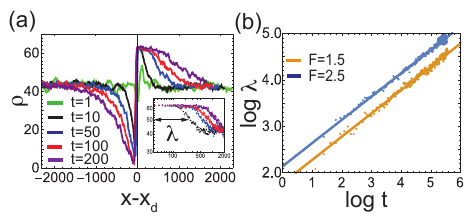}
\caption{(a) Density of the passive particles $\rho$ as function of the position, relative to the center of mass of the driven particles. An increasing length scale $\lambda$ occurs which signifies the size of the jammed front. (b) The time evolution of $\lambda$, confirming the $\lambda \sim \sqrt{t}$ scaling. The same setup was used as in Fig. \ref{fig:3}.  \label{fig:2}}
\end{figure}

\paragraph{Formation and growth of the jammed front.} 
Although the driven particles no longer evolve beyond the single cluster state, the presence of this single cluster becomes increasingly dominant for the behavior of the entire system. As the cluster progresses through the system, it continues to sweep up passive particles. The passive particles in front of the driven particles become jammed and fully immobilized. This jammed zone, in turn, acts as a plunger propagating through the system. To explore the formation and growth of this jammed front, we turn to larger system sizes. This ensures the system has sufficient space and time to develop, and that there is no direct interaction between the compressed zone, and the wake behind the cluster (see also Fig. \ref{fig:3}). We distribute 50 particles randomly in the subdomain $-50 < z < 50$, and immersed them in 19950 passive particles which are randomly distributed in the full system with size $-2320 < z < 2320$. The system width remains 6. A section of the full system is shown in Fig. \ref{fig:3} (a)). This large system allows us to study the formation of the jammed front, eliminating the effects of the lateral boundaries. First, we report the time evolution of the order parameter $\sigma$ for various driving force $F_d$ (See Fig. \ref{fig:3} (a)). Initially, the driven particles move individually through the passive particles until, at some point, the region in front of the driven particles is so densified that it becomes jammed. This prompts the formation of the cluster of driven particles. Subsequently, the region with elevated density in front of the cluster increases in size $\lambda$ (See Fig. \ref{fig:2} (a)). That the compressed zone should grow over time is to be expected: since the jammed particles are no longer able to move by diffusive motion, they act in unison with the driven cluster, and - even though they are not themselves driven - continue to sweep up nondriven particles in front. The evolution of the width of this front can thus be understood by means of a simple argument. Assuming that the cluster, and the jammed zone, move at some velocity $v_f$, and that this velocity results from a driving force $F_d$. This defines an effective friction coefficient $\xi=v_f/F_d$. The coefficient $\xi$ should scale with the number of particles $n$ that the front is pushing. On the other hand, a front moving at velocity $v_f$ sweeps up a number $\Delta n \sim v_f \Delta t$ per time interval $\Delta t$, and these are added to the front. Thus we have two relations for the velocity: $v_f \sim F_d/n$ and $v_f \sim {\rm d} n/{\rm d} t$. From this, we see that ${\rm d} n/{\rm d} t\sim F_d/n$ which means that $n(t)\sim \sqrt{F_d t}$. Since the height is constant, this result means that the width of the compressed zone should scale as $t^{1/2}$. Fig. \ref{fig:2} (b) confirms that indeed, for various forces, the size of the jammed front grows as the square root of time.

\paragraph{Dynamics of the driven particles.} 
Driving in dense dispersions is typically applied to effect transport. A natural question to ask is how efficiently the driven particles may be transported in this manner. Does clustering help, or hinder, their passage through the background phase? When we measure the velocity $v_d$ of the driven particles upon clustering we find that the average velocity at which the individual particles move through the passive particles is higher than when the particles have clustered. So, in general, fully clustered states are non-optimal for transport. Clearly, however, no driving at all is also not optimal as then, no transport occurs. The question then arises: is there an {\em optimal} driving force for the transport of driven particles. We now measure the time-evolution of the center of mass of the driven particles, at a fixed density $\phi$ and various forces $F_d$. Fig. \ref{fig:3} (c) shows that the velocity evolves with time very differently before and after the clustering. Before clustering, the velocity is constant confirming an effective friction coefficient that is independent of time. After clustering occurs, the velocity drops to a different curve and scales as $t^{-1/2}$, pointing to an effective friction coefficient that grows as the square root of time - in agreement with the prediction from the argument we presented before.  The time at which the cluster forms depends inversely on $F_d$ (See Fig. \ref{fig:3} (d)). This permits us to extract the optimal force for transport: transport of the driven particles is most effective at a force just prior to clustering, where they still travel efficiently through the medium but do not perturb it sufficiently to cluster and compress it (See Fig. \ref{fig:3} (e)). 

\paragraph{Conclusions}

\begin{figure}
\centering
\includegraphics[scale=2]{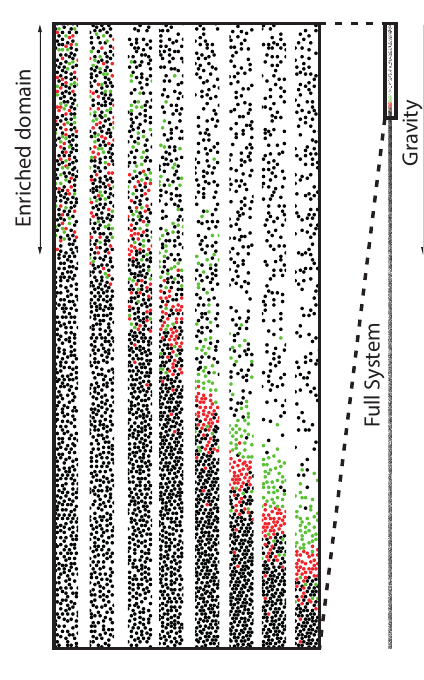}
\caption{(left) Sedimentation of particles with mass 0.5 (green) and 2, randomly distributed in a subdomain $100< z < 950$, where a reflecting wall is placed at the top and bottom of the system ($T=1$, $l=1000$ and $w=6$)  \label{fig:sed}}
\end{figure}

To summarize, we report that, depending on the driving force $F_d$, the density $\phi$, the composition $N_d/N_c$ and the temperature $T$ a dynamically induced clustering occurs in mixtures of driven and passive particles. A late times, the resulting system-spanning cluster dominates the global transport properties of the mixture. This effect is generic, and we hypothesize it to occur in a wide range of differentially driven, dense systems. One specific application that we foresee is to use the clustering mechanism to separate colloidal particles by mass. Fig. \ref{fig:sed} shows the basic idea and operating principle: a random mixture of light (green) and heavy (red) particles is poured into the top of a sedimentation column filled with density-matched colloids. Since both the light and the heavy particles are driven differently from the background, both phase separate into tight clusters. The cluster itself, however, is also inhomogeneous in the sense that very quickly, the light particles are at the back and the heavy ones in front. This way, gravity can dynamically phase separate the various populations of particles by mass: {\em colloidal chromatography}. Whether or not clustering will occur in this case will similarly depend on density, mass and composition. Preliminary simulations show that the mass separation continues to work for much broader polydisperse mass distributions, and in an upcoming publication we explore the concept of the colloidal chromatograph in much greater detail. 

\begin{acknowledgments}
We thank Stefan Paquay for helpful discussions. This work was supported by funds from the Netherlands Organization for Scientific Research (NWO-FOM) within the programme "Barriers in the Brain: the Molecular Physics of Learning and Memory" (No. FOM-E1012M).
\end{acknowledgments}

\bibliography{biblio}

\end{document}